\documentclass[aps,prb,twocolumn]{revtex4}
\usepackage{amssymb,lipsum}
\usepackage{graphicx,amsmath}
\setcitestyle{numbers,square}

\newcommand{\bea}{\begin{eqnarray}}
\newcommand{\eea}{\end{eqnarray}}
\newcommand{\beq}{\begin{equation}}
\newcommand{\eeq}{\end{equation}}
\newcommand{\benu}{\begin{enumerate}}
\newcommand{\enu}{\end{enumerate}}

\newcommand{\si}{\sigma}

\newcommand{\bk}{{\bf k}}
\newcommand{\bq}{{\bf q}}

\newcommand{\br}{{\bf r}}

\begin{document}

\title{Pairing instability near a lattice-influenced nematic quantum critical point}
\date{\today}
\author{D. Labat$^1$ and I. Paul$^1$}
\affiliation{
$^1$Laboratoire Mat\'{e}riaux et Ph\'{e}nom\`{e}nes Quantiques, Universit\'{e} Paris Diderot-Paris 7 \& CNRS,
UMR 7162, 75205 Paris, France
}

\begin{abstract}

We study how superconducting $T_c$  is affected as an electronic system in a tetragonal environment is tuned to a
nematic quantum critical point (QCP). Including coupling of the electronic nematic variable to the relevant lattice
strain restricts criticality only to certain high symmetry directions. This allows a weak-coupling treatment, even
at the QCP. We develop a criterion distinguishing weak and strong $T_c$ enhancements upon approaching the QCP. We
show that negligible $T_c$ enhancement occurs only if pairing is dominated by a non-nematic interaction away from
the QCP, and simultaneously if the electron-strain coupling is sufficiently strong. We argue this is the case of the
iron superconductors.

\end{abstract}


\maketitle

\section{Introduction}
The origin of high superconducting transition temperature $T_c$ of the copper and iron based systems
remains to be well understood~\cite{bednorz86,kamihara08,anderson87,review-cuprate,review-FeAs}.
Among the various possibilities
as likely causes of high $T_c$, one is that of the presence of a quantum critical point (QCP) in
the vicinity of which the effective pairing interaction is strong, leading to enhanced $T_c$. In fact,
$T_c$ boosted by an antiferromagnetic QCP remains among the more promising scenarios for
these systems~\cite{afm-qcp}. A related question, addressed here, is whether one expects similar enhancement
of $T_c$ close to a nematic QCP,
where the ground state is poised to break discrete rotational symmetry.

Several recent theoretical works on this issue have concluded that, indeed, the superconducting
$T_c$ is boosted near a nematic QCP~\cite{yamase2013,maier2014,lederer2015,metlitski2015}.
The motivation for this conclusion is intuitively clear, at least in the
regime where weak-coupling theory of pairing is applicable. It is well-known that, for small momentum transfer,
the effective electron-electron interaction mediated by the long-wavelength nematic fluctuations
leads to attractive interaction both in the $s$- and $d$-wave channels. This interaction
increases as the system approaches a nematic QCP, leading to larger $T_c$.
Refs.~\onlinecite{yamase2013} and~\onlinecite{metlitski2015} have studied the problem beyond weak coupling, and both
have concluded that the intuitive picture stays intact.

In the phase diagram of the cuprates the location of a nematic QCP, if present, remains to be well established.
Consequently, at present the Fe-based systems are better suited to study the issue of $T_c$ enhancement from an experimental
point of view. However, while for most iron based superconductors (FeSC) the nematic transition line and the QCP is
well-identified~\cite{chu2012,Gallais2013,Fernandes2014,Gallais-2016-prl,Gallais2016,kuo2016},
the presence of a magnetic transition and its associated QCP complicates matter~\cite{review-FeAs},
since it is hard to disentangle the effects of the two.
In this respect an ideal system is FeSe$_{1-x}$S$_x$, which has a nematic QCP but not a magnetic one. Interestingly,
in contradiction with the above theoretical expectations,
in this model system the $T_c$ is hardly affected by the nematic QCP~\cite{watson15,hosoi16,urata16}.
Thus, clearly there is a missing element in the above theories.

In this work we identify the missing element to be
a symmetry-allowed coupling between the electronic nematic degree of freedom and a lattice shear strain mode.
We show that, once this coupling is included in the theory of nematic criticality, the presence of a nematic QCP
does not necessarily lead to significant enhancement of $T_c$. We identify the conditions under which the
enhancement is negligible near the QCP.
This occurs when the following two conditions are simultaneously satisfied. Namely, (i)
the pairing is  dominated by a non-nematic interaction \emph{away} from the QCP, and (ii)
if the electron-strain coupling is sufficiently large.
Note, condition (i) does not trivialize the problem since, by itself,
it does not preclude the nematic term to dominate \emph{near} the QCP
and provide significant $T_c$ enhancement.
Our result provides a route
to understand qualitatively why $T_c$ is unaffected by the nematic QCP in FeSe$_{1-x}$S$_x$.

The main physics ingredient of our work is enshrined in the standard theory of elasticity for an acoustic
instability involving Ising-nematic symmetry, such as a second order
tetragonal-orthorhombic transition. It is well known that, in this case the divergence of the correlation length,
which manifests as vanishing acoustic phonon velocity, \emph{is restricted to two high-symmetry
directions}~\cite{Landau-Lifshitz,Larkin1969,Levanyuk1970,Cowley1976,Folk1976,Zacharias2015}.
This is because along the generic directions the non-critical shear
strains, that are invariably present in a solid, come into play and cutoff criticality. The physics
of this cutoff can be also understood as follows.

Consider a translation symmetry preserving second order phase transition involving a local variable
$\mathcal{X}(\br) = \mathcal{X}_0 + \sum_{\bq \neq 0} \mathcal{X}_q e^{i \bq \cdot \br}$, where $\mathcal{X}_0$ is the
order parameter.
Within the Landau paradigm the free energy has mean field and fluctuation contributions.
The former has the structure $F_{\rm MF} = (a/2) \mathcal{X}_0^2 + (A/4) \mathcal{X}_0^4$, while the latter, to Gaussian order,
is related to the action $S_{\rm fluc} = \sum_{\bq \neq 0} (b + q^2) \left| \mathcal{X}_q \right|^2$.
While in usual theories $b=a$, in those involving crystalline strains $b(\hat{q})$ is no longer a parameter but, rather, a
function of the Brillouin zone angles $\hat{q}$ containing information about the
crystalline anisotropy~\cite{Larkin1969,Levanyuk1970}.
In other words, the concept of correlation length becomes angle-dependent.
In this situation, the condition $b(\hat{q}) =a$ is satisfied only for certain high symmetry directions, and only along
these directions the correlation length diverges at the transition defined by $a=0$.
Along other directions $b(\hat{q}) >0$, and correlation length stays
finite at the transition.

\begin{figure}[!!t]
\begin{center}
\includegraphics[width=1.0\linewidth,trim=0 0 0 0]{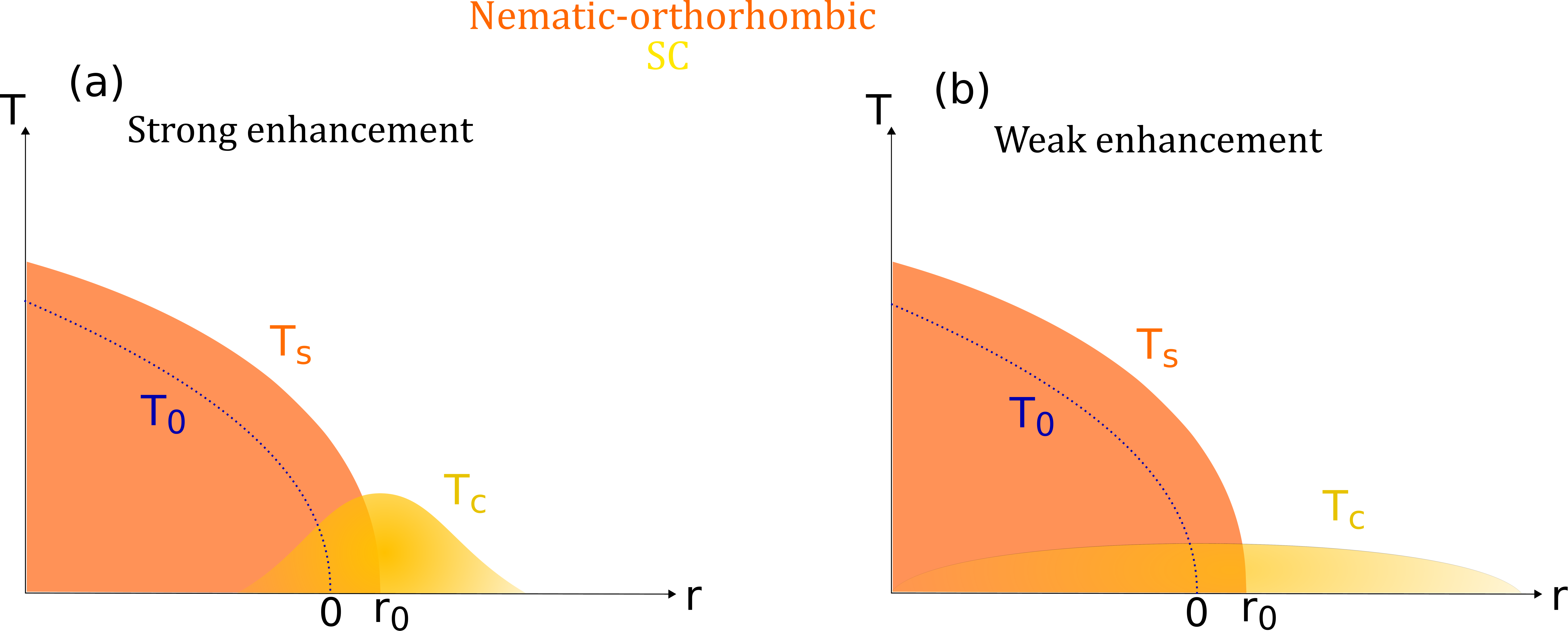}
\caption{
(color online) Schematic phase diagrams with nematic/orthorhombic (orange) and superconducting (yellow) phases
below temperatures $T_s(r)$ and $T_c(r)$, respectively. $r$ is a dimensionless control parameter.
$T_0(r)$ (dotted-lines) is the nominal electron-nematic transition in the absence of nemato-elastic coupling.
The coupling shifts the nematic quantum critical point (QCP) from $r=0$ to $r=r_0$.
$r_0$ is a measure of the strength of the coupling (see text).
(a) and (b) are two possible scenarios. In (a) there is ``strong'' enhancement of $T_c(r)$ at the QCP. In (b) the enhancement is
``weak'', as in FeSe$_{1-x}$S$_x$~\cite{watson15,hosoi16,urata16}.
We show that (b) occurs if the pairing is dominated by a non-nematic potential away from the QCP, and simultaneously
if the nemato-elastic coupling is sufficiently strong (see Eq.~(\ref{eq:condition})).
}
\label{fig1}
\end{center}
\end{figure}

The above property is inherited by the
electronic nematic subsystem once its coupling with the strain is included~\cite{Cano2010,Karahasanovic2016,paul2016}.
This leads to
two important conclusions concerning Cooper pairing, which are the main results of this paper.
(i) Under certain standard assumptions, the weak coupling BCS analysis remains valid arbitrarily close to the nematic QCP.
(ii) We identify the criterion that distinguishes between strong and weak $T_c$ enhancements
upon tuning the system to the nematic QCP (see Fig.~\ref{fig1}). The latter occurs only if
the pairing is dominated by a non-nematic interaction away from the QCP, and simultaneously
if the electron-phonon interaction is sufficiently strong. We argue that this is the case of the FeSC.

\section{Model}
We consider a system of itinerant electrons in a tetragonal lattice, with negligible dispersion along
the $z$-axis, which is close to a nematic/structural QCP that is driven by electronic correlations.
Ignoring electron-lattice interaction for the moment, the long wavelength fluctuations of the nematic variable
$\phi_{\bq}$, which is a collective mode of the electrons, is described by a susceptibility of the standard Ornstein-Zernike
form $\chi_0^{-1}(\bq) = r + q_{2d}^2/(2k_F)^2$, where $\bq_{2d} \equiv (q_x, q_y)$,
and $r$ is a dimensionless tuning parameter of the theory that governs closeness to the QCP.
Without the lattice coupling, the bare QCP is at $r=0$. In what follows the frequency dependence of the susceptibility
can be ignored.

A crucial ingredient in the model is the symmetry-allowed nemato-elastic term linking
$\phi (\br)$ with the local orthorhombic strain
$\varepsilon({\bf r}) = \varepsilon + i \sum_{\bq \neq 0}
\left[ q_x u_x(\bq) - q_y u_y(\bq) \right] e^{i \bq \cdot \br}$, where
$\varepsilon$ is the uniform macroscopic strain, and
$\vec u({\bf r})$ is the atomic displacement. $\varepsilon$ is
non-zero only in the symmetry-broken nematic/orthorhombic phase.
This coupling can be written as
$
g \int d{\bf r} \phi (\bf {r}) \varepsilon(\bf {r}),
$
where $g$ has dimension of energy.
\begin{figure}[!!t]
\begin{center}
\includegraphics[width=1.0\linewidth,trim=0 0 0 0]{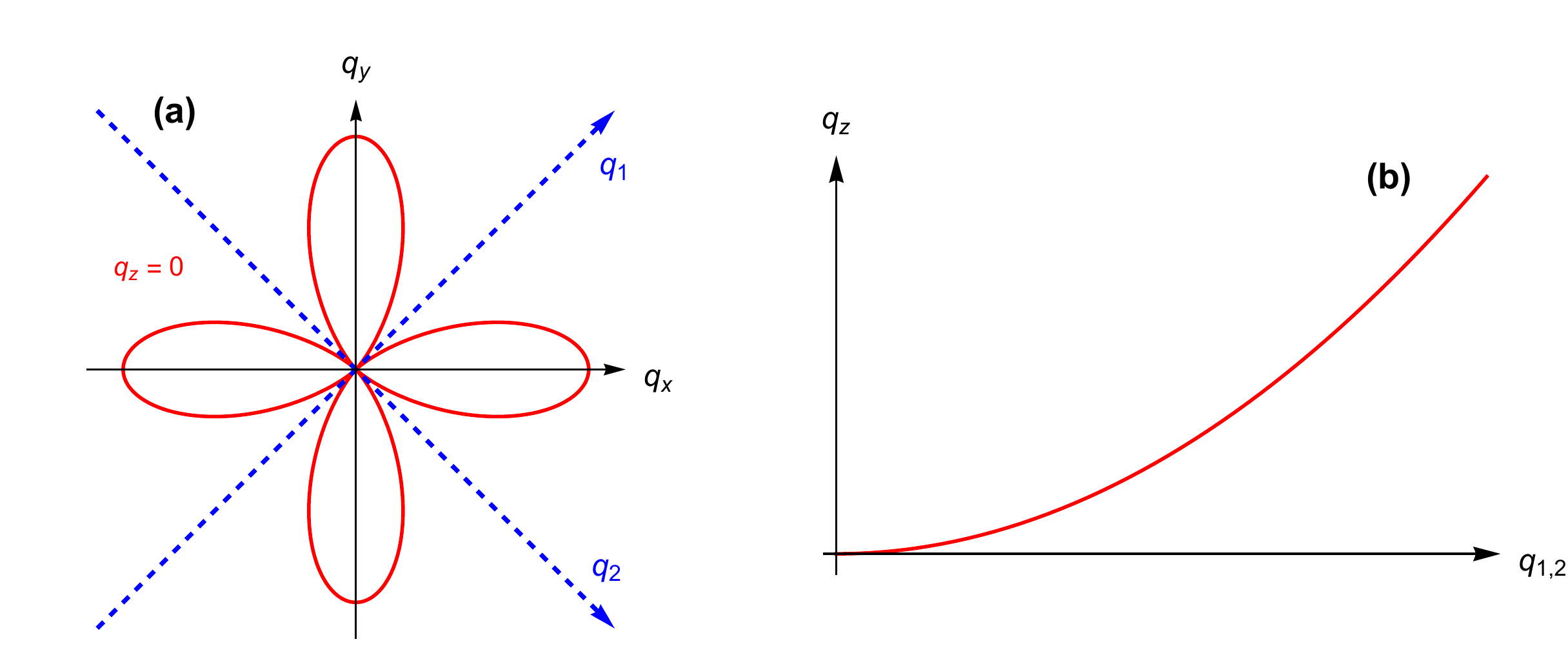}
\caption{
(color online) The mass of the nematic susceptibility $\chi$, defined as
$r(\hat{q}) \equiv \lim_{\bq \rightarrow 0} \chi^{-1} (\bq, \omega =0)$ becomes anisotropic in the presence of the
nemato-elastic coupling. In the non-nematic phase the angular dependence of $r(\hat{q})$ has tetragonal symmetry.
(a) Variation of $r(\hat{q})$ on the $q_z=0$ plane (red line) at the nematic QCP . It is zero only along the high-symmetry directions
$\hat{q}_{1,2} \equiv (\hat{q}_x \pm \hat{q}_y)/\sqrt{2}$ (blue arrows). (b) For finite $q_z$, $r(\hat{q}) \propto q_z^2$.
}
\label{fig2}
\end{center}
\end{figure}

\begin{figure}[!!t]
\begin{center}
\includegraphics[width=1.0\linewidth,trim=0 0 0 0]{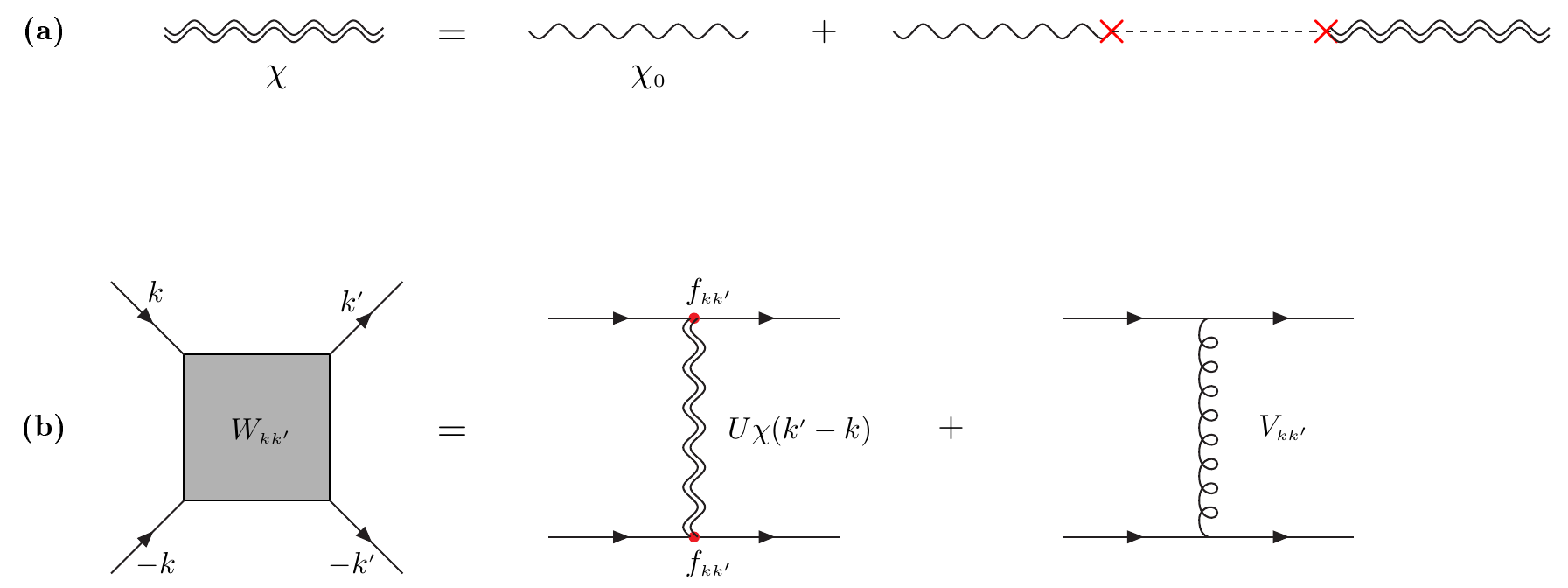}
\caption{
(color online) Diagrammatic representation of the relevant microscopic processes.
(a) The bare electron-nematic susceptibility (single wavy-line) is dressed (double wavy-line) by the
nemato-elastic coupling (red crosses). The dashed line is an acoustic phonon.
(b) The pairing potential $W_{\bk, \bk^{\prime}}$ consists of the dressed nematic interaction of strength
$U$, and a non-nematic interaction $V_{\bk, \bk^{\prime}}$ (gluon-line) of strength $V$. The former interaction
vertex is accompanied by a form factor $f_{\bk, \bk^{\prime}}$. The interesting regime is $V > U$, i.e., when
pairing is dominated by the non-nematic interaction away from the nematic QCP.
}
\label{fig3}
\end{center}
\end{figure}

The effect of the nemato-elastic term on criticality has been discussed earlier~\cite{Cano2010,Karahasanovic2016,paul2016}.
Here, for the sake of completeness, we recapitulate the main points.
(i) It shifts the QCP
to $r = r_0  \equiv g^2 \nu/C_0$ (see Fig.~\ref{fig1}), where $\nu$ has dimension of density of states
and $C_0$ is the bare orthorhombic elastic constant.
Thus, $r_0$ is a dimensionless parameter that measures the strength of the nemato-elastic coupling.
In the following we take $r_0 \leq r \leq 1$.
(ii) The nemato-elastic coupling leads to
hybridization of $\phi_{\bq}$ with the acoustic phonons (see Fig.~\ref{fig3}(a)), which renormalizes
the nematic susceptibility to
$\chi^{-1} = \chi^{-1}_0 - \Pi$, with
$
\Pi(\hat{q}) = (g^2/\rho) \sum_{\mu}
\left( {\bf a}_{\bq} \cdot \hat u_{\bq, \mu} \right)^2/
\omega^2_{\bq, \mu}.
$
Here $\rho$ is the density, $\mu$ is the
polarization index, $ {\bf a}_{\bq} \equiv (q_x, -q_y,0)$,
and $\hat u_{\bq, \mu}$
is the polarization vector for the bare acoustic phonons
with angle-dependent velocity ${\bf v}^{(0)}_{\hat{q},\mu}$ and dispersion
$\omega_{\bq, \mu} = {\bf v}^{(0)}_{\hat{q},\mu} \cdot \bq$.
The above follows simply from integrating out the lattice variables.
Evidently, $\Pi(\hat{q})$ is independent of the magnitude $q$, and has four-fold
symmetry of the tetragonal unit cell in the non-nematic phase.
Thus, the nemato-elastic term makes the mass of $\phi_{\bq}$ angle-dependent with
$r \rightarrow r(\hat{q}) \equiv r + \Pi(\hat{q})$, and criticality, or divergence of
correlation length, is restricted to the high symmetry directions
$\hat{q}_{1,2} \equiv (\hat{q}_x \pm \hat{q}_y)/\sqrt{2}$, for which $r(\hat{q}_{1,2}) =0$
at the QCP (see Fig.~\ref{fig2}). Along the remaining directions $r(\hat{q}) > 0$ at the QCP.
Note, since divergence of $\chi$ also implies vanishing of the sound velocity \emph{renormalized}
by the coupling $g$,
the above direction dependence is consistent with the fact that only along $\hat{q}_{1,2}$ the sound
velocity vanishes at this nematic/structural transition~\cite{yoshizawa2012,boehmer2014}.
(iii) Since the coupling $g$ cuts off divergence along the generic directions, the effect of the
quantum fluctuations is weak. Therefore, below the temperature scale $T_{\rm FL} \sim r_0^{3/2} T_F$
the system behaves as a Fermi liquid for thermodynamic and single-electron properties~\cite{paul2016}. Here $T_F$
is the Fermi temperature. Above $T_{\rm FL}$ the nemato-elastic coupling can be neglected.

The effective electron-electron interaction mediated by the nematic variable $\phi_{\bq}$
has the form
\beq
\label{eq:H-nem}
\mathcal{H}_{\rm nem} = - U \sum_{\bq} \chi(\bq) \mathcal{N}_{\bq} \mathcal{N}_{-\bq},
\eeq
where $\mathcal{N}_{\bq} = \sum_{\bk, \si} f_{\bk,\bq} c^{\dagger}_{\bk + \bq, \si} c_{\bk, \si}$, in terms of the
creation and annihilation operators of electron with spin $\si$. The form factor
$f_{\bk,\bq} = (h_{\bk} + h_{\bk + \bq})/2$, where $h_{\bk}$ transforms as $(k_x^2 - k_y^2)$. The parameter $U$, with dimension of
energy, sets the scale of the nematic interaction.

In the following we solve the linearized BCS equations for superconducting gap $\Delta_{\bk}$, assuming singlet pairing.
This involves calculating the largest eigenvalue $\lambda$ satisfying the relation
\beq
\label{eq:BCS}
\lambda \Delta_{\bk} =  \nu_{\rm FS} \oint_{{\rm FS}^{\prime}} W_{\bk, \bk^{\prime}} \Delta_{\bk^{\prime}},
\eeq
where $\nu_{\rm FS}$ is the density of states at the Fermi surface,
\beq
\label{eq:kernel}
W_{\bk, \bk^{\prime}}  = V^{(s,d)}_{\bk, \bk^{\prime}}
+ U  f_{\bk, \bk^{\prime}}^2 \chi(\bk - \bk^{\prime}),
\eeq
and ${\rm FS}^{\prime}$ implies the $\bk^{\prime}$-integral is restricted to the Fermi surface (see Fig.~\ref{fig3}(b)).

In the above we added
a second interaction $-V^{(s,d)}$ of non-nematic origin that stays constant as a function of $r$.
Depending on the context, it favors $s$- and $d$-wave pairing, respectively.
The addition of $V^{(s,d)}$ can be motivated as follows.
In all likelihood, the pairing in the FeSC and the cuprates is mediated not just by
the nematic fluctuations. In addition, there is, e.g., short wavelength spin/charge fluctuations~\cite{review-FeAs}
or Mott correlations~\cite{anderson87,review-cuprate} that mediate pairing.
Consequently, it is physical to expect that, close to a nematic QCP, the pairing potential has a nematic component
which is a strong function of $r$ (included in $\chi(\bq)$),
and it has a non-nematic component which does not vary with $r$ (represented by $V^{(s,d)}$).
Note, in what follows the precise microscopic origin and structure of $V^{(s,d)}$ is not relevant.
Besides this physical relevance, as we show below, the inclusion of $V^{(s,d)}$ is crucial
to distinguish the two limiting cases of ``strong'' and ``weak'' enhancement of $T_c$ upon tuning
the system to the nematic QCP with $r \rightarrow r_0$.

Our goal is to study how $\lambda(r)$ changes as the system is tuned to the QCP, from which we can deduce
the variation of $T_c \sim \Lambda e^{-1/\lambda}$, where $\Lambda \ll E_F$ is the high-energy cutoff of the pairing problem.
Note, an \emph{important consequence} of the coupling $g$ is that, it is now possible to consider the case where
$T_c(r_0) < T_{\rm FL}$, the Fermi liquid scale.
For $T < T_{\rm FL}$ the dynamics of the nematic pairing potential is irrelevant, and the problem can be treated within
BCS formalism.

In the above model $\lambda(r)$ increases monotonically as the system approaches the QCP, since the
nematic interaction itself is attractive and monotonic.
However, the crucial question is whether this increment is significant. To address this issue
quantitatively, we define $\delta \lambda \equiv \lambda(r=r_0) - \lambda(r=1)$, and we distinguish
between ``strong'' and ``weak'' enhancements of $T_c$, depending on whether $\delta \lambda \gg \lambda(r=1)$ or not,
respectively. Qualitatively, this criteria distinguishes between whether pairing is dominated by long wavelength
nematic fluctuations or by a non nematic pairing interaction at the QCP.

\section{Results}
The momentum anisotropy of the susceptibility $\chi(\bq)$ due to the coupling $g$
can be modeled as follows. (a) For $q_z \leq q_{2d}$,
we get $\chi^{-1}(\bq)\approx r(\hat{q}) + q_{2d}^2/(2k_F)^2$. The anisotropic mass
$r(\hat{q})$ has tetragonal symmetry, and satisfies $r(\hat{q}_{1,2}) = 0$ at the QCP. The simplest function consistent
with these requirements is
$r(\hat{q}) = (r-r_0) + r_0 (q_z/q_{2d})^2 + r_0 \cos^2 2 \phi_{\bq}$, where
$\phi_{\bq}$ is the azimuthal angle of $\bq$ (see Fig.~\ref{fig2}). This region of $\bq$-space also contains the critical modes.
(b) For  $q_z \geq q_{2d}$, the nemato-elastic coupling can be neglected and
$\chi(\bq) \approx \chi^{-1}_0(\bq) = r + q_{2d}^2/(2k_F)^2$.
However, this does not imply singular susceptibility at the QCP, since its location is shifted from
$r=0$ to $r=r_0$. In this region of $\bq$-space the modes are, thus, non-critical.

The main qualitative physics can be already illustrated by considering the simplest case of a
single band with a cylindrical Fermi surface around the Brillouin zone center, and where the non-critical
pairing term supports $s$-wave gap with $V^{(s)}_{\bk, \bk^{\prime}} = V >0$. The details of the calculation are given in the
Appendix. For a uniform gap the leading $r$-dependence of the eigenvalue $\lambda(r \geq r_0)$ is given by
\beq
\label{eq:result}
\lambda/\nu_{\rm FS}
= V  + \frac{U}{2 \sqrt{r}} - \frac{U}{\pi} \left( \ln {\rm max}[r-r_0,r_0] + c_1 \right),
\eeq
where $c_1 = 8/3 - 2 \ln 2 \approx 1.28$ is non-universal.
In r.h.s of the above the second term comes from the momentum space (b), as discussed above, where the fluctuations
are massive and non-critical.
While the third term comes from the region (a) which includes
the critical modes. However, since the critical momentum space is rather restricted (equivalently, the critical
theory can be mapped to an isotropic model in effective space dimension $d_{\rm eff} =5$~\cite{Folk1976}), its contribution to
the eigenvalue is subleading. Thus, the leading nematic contribution to $\lambda$ is from the non-critical region (b).
\begin{figure}[!!t]
\begin{center}
\includegraphics[width=1.0\linewidth,trim=0 0 0 0]{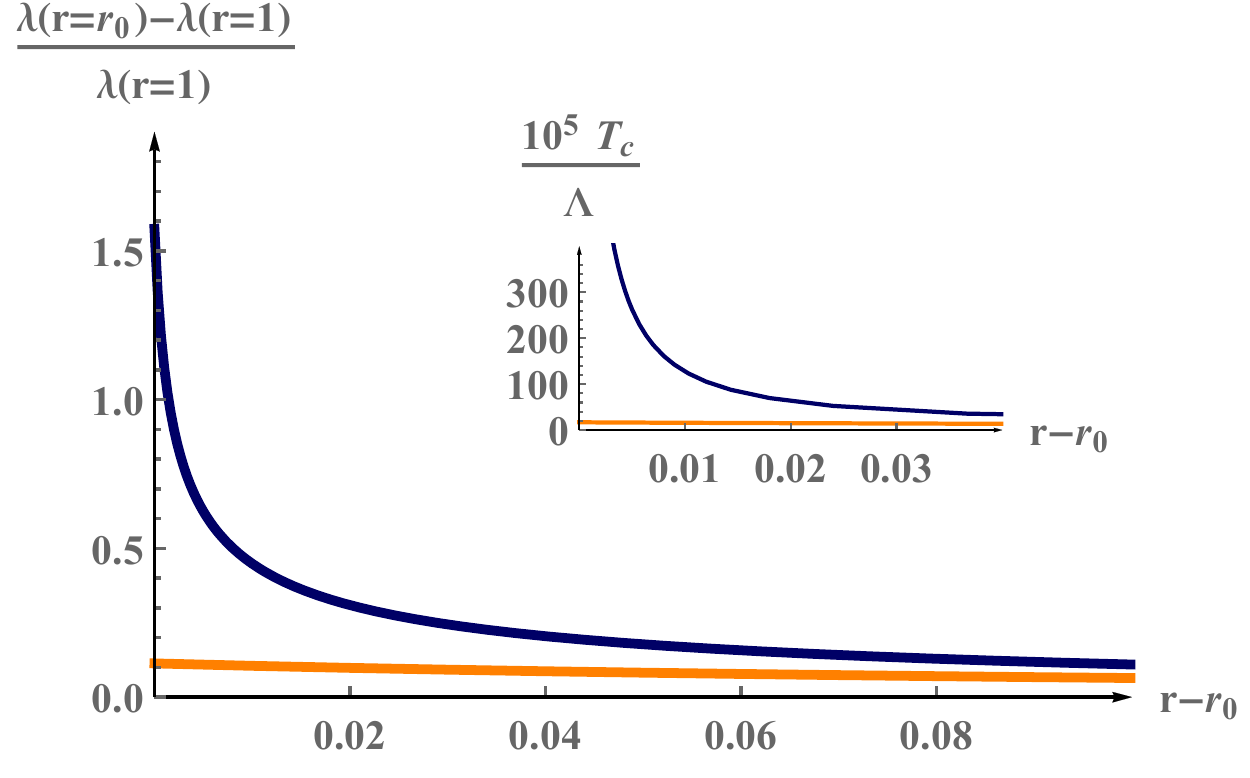}
\caption{
(color online) Calculated variation of the BCS eigenvalue $\lambda(r)$ (main panel)
and the associated superconducting transition temperature $T_c(r)$ (inset) upon tuning the
system to the nematic QCP at $r=r_0$. The pairing kernel is shown graphically in Fig.~\ref{fig3}.
Large enhancement (blue curves) of $\lambda$ and $T_c$ is observed only when the nemato-elastic
coupling strength $r_0$ is sufficiently weak (see Eq.~(\ref{eq:condition}). Note, in this limit, at the QCP the
$T_c \propto e^{-\sqrt{r_0}/(\nu_{\rm FS} U)}$ (not shown) is still a small fraction of the ultraviolet cutoff $\Lambda$.
In the opposite limit of strong nemato-elastic coupling (orange curves) the enhancement is negligible. This limit is
relevant for the Fe-based superconductors such as FeSe$_{1-x}$S$_x$~\cite{watson15,hosoi16,urata16}.
}
\label{fig4}
\end{center}
\end{figure}

Importantly, from the above we deduce that, upon tuning the system to the QCP by $r \rightarrow r_0$, the $T_c$ enhancement
will be ``significant'', i.e., $\delta \lambda \gg \lambda(r=1)$, provided the electron-lattice coupling is weak enough
such that
\beq
\label{eq:condition}
r_0 \ll (U/V)^2.
\eeq
Note, if the nemato-elastic coupling is ignored we get $r_0=0$, and we recover the result
of Refs.~\cite{yamase2013,maier2014,lederer2015,metlitski2015} namely, the presence of a nematic QCP
necessarily implies strong $T_c$ increase.
Note also, the complete absence of a non-nematic pairing term ($V=0$) leads to strong $T_c$ enhancement as well.

As importantly, the opposite limit of weak $T_c$ enhancement, which is relevant for understanding the phase diagram of
FeSe$_{1-x}$S$_x$, occurs if $1 \gg r_0 \gg (U/V)^2$ (see Fig.~\ref{fig4}).
Physically, this implies a situation where the following two conditions are simultaneously satisfied.
(i)The pairing is  dominated by a non-nematic interaction \emph{away} from the QCP (since $V \gg U$), and (ii)
the electron-strain coupling is sufficiently large (since $r_0 \gg (U/V)^2$).
Note, condition (i) does not trivialize the issue since, by itself,
it does not preclude the nematic term to dominate \emph{near} the QCP
and provide significant $T_c$ enhancement. In fact, this is why condition (ii) comes into play. The origin of condition
(i) lies in the physical expectation that the energy scale generated by the electron-lattice interaction is well below the
Fermi energy, i.e., $r_0 \ll 1$.

Besides the case of the isotropic $s$-wave gap, we also study the following situations in the Appendix.
(i) An extended $s$-wave gap, since the lattice-renormalized nematic interaction is intrinsically anisotropic,
and it can give rise to angular variations of the gap. (ii) Motivated by the cuprates, we consider the case where
the non-nematic interaction $V^{(d)}_{\bk, \bk^{\prime}} = V \cos 2 \phi_{\bk} \cos 2 \phi_{\bk^{\prime}}$
favors a $d$-wave gap with $\Delta_{\bk} = \Delta_0 \cos 2 \phi_{\bk}$. (iii) Motivated by the FeSC we study a
system with Fermi pockets at $(0,0)$, $(\pm \pi,0)$ and $(0, \pm \pi)$, with a form of $V^{(s)}_{\bk, \bk^{\prime}}$ that
leads to $s_{\pm}$-gap. In all these cases we find that qualitatively $\lambda(r)$ is described by Eq.~(\ref{eq:result}),
except with different numerical pre-factors. We conclude
that the above criterion for $T_c$ enhancement in Eq.~(\ref{eq:condition}) is robust.

\section{Conclusion}
The strength of the nemato-elastic coupling can be estimated
as $r_0 \sim (T_s - T_0)/T_F$, where $T_0$ is the nominal nematic transition temperature of
the electron subsystem in the absence of this coupling (see Fig.~\ref{fig1}), accessible
from, say, electronic Raman scattering~\cite{Gallais2016}.
In FeSe we get $T_0 \sim$ 10 K, and $T_s \sim$ 90 K~\cite{massat2016}. We estimate the Fermi temperature from the bottom of the smallest
electron pocket as measured by photoemission above $T_s$, which is around 25 meV in FeSe~\cite{watson15,fanfarillo2016}.
Thus, for FeSe$_{1-x}$S$_x$ we estimate
$r_0 \sim 0.3$ and $T_{\rm FL} \sim 40$ K. Note, the condition $T_c < T_{\rm FL}$, needed for a weak-coupling theory, is well-respected in this case.
A similar estimate for Ba$($Fe$_{1-x}$Co$_x)_2$As$_2$ yields $r_0 \sim$ 0.05 and $T_{\rm FL} \sim 10$ K~\cite{paul2016}.
Since, in this system the maximum $T_c \sim 25$ K  is comparable to $T_{\rm FL}$, a more careful quantitative analysis is needed.

The estimation of $U$ and $V$ requires a full microscopic theory of pairing, that is currently unavailable.
Consequently, a quantitative application of the theory to real systems is not possible at present.
However, experimentally it is clear that FeSe$_{1-x}$S$_x$ has a nematic
QCP around $x \approx$ 0.16, but the superconducting $T_c(x)$ remains remarkably flat around this
doping~\cite{watson15,hosoi16,urata16}. This can be due to strong nemato-elastic effect violating
the condition in Eq.~(\ref{eq:condition}). In turn, this would imply that the
pairing interaction in FeSe$_{1-x}$S$_x$ is mostly non-nematic in origin.
A similar case can also be made for Ba$($Fe$_{1-x}$Co$_x)_2$As$_2$, where quantum critical nematic fluctuations have been
detected only over a narrow doping range of $x=$ 0.65 - 0.75 in the low-$T$ superconducting phase~\cite{Gallais-2016-prl}.
It is remarkable that
over the same doping range $T_c(x)$ hardly varies, implying that even here the lattice cutoff is operational.
Thus, the dome-like structure of $T_c(x)$ over a wider doping range in Ba$($Fe$_{1-x}$Co$_x)_2$As$_2$ is
likely due to the antiferromagnetic QCP, while the absence of a magnetic QCP in FeSe$_{1-x}$S$_x$ results in a flat $T_c(x)$.

To summarize, we argued that nemato-elastic coupling can play a crucial role
in determining if superconducting $T_c$
is strongly enhanced in the vicinity of a nematic quantum critical point. We showed that, in the presence of a
significant non-nematic pairing interaction, strong
nemato-elastic coupling implies that the nematic fluctuations do not boost $T_c$ significantly. Based on
existing experiments
on FeSe$_{1-x}$S$_x$ and on Ba$($Fe$_{1-x}$Co$_x)_2$As$_2$ we argued that this is likely the case of the
iron-based superconductors. This would imply that the main pairing interaction is non-nematic in origin
in these materials. More generally, from the perspective
of material design for high temperature superconductivity, we conclude that (a) hard crystals are
better suited for boosting $T_c$ near
a nematic quantum critical point, and that (b) the lattice cutoff can be also avoided provided the
non-nematic pairing potential
is strong enough to guarantee $T_c (r=1) \gg T_{FL}$, in which case the
physics of Refs.~\cite{yamase2013,maier2014,lederer2015,metlitski2015} will be operational.

\acknowledgments
We are thankful to M. Civelli, Y. Gallais and M. Garst for insightful discussions.
I.P. acknowledges financial support from ANR grant ``IRONIC'' (ANR-15-CE30-0025).

\appendix
\section{}

In this appendix we provide the technical details for the calculation of the BCS eigenvalue $\lambda$ defined
in Eq.(~\ref{eq:BCS}) of the main text which is
\begin{equation}
\lambda \Delta_{\bk} =  \nu_{\rm FS} \oint_{{\rm FS}^{\prime}} W_{\bk, \bk^{\prime}} \Delta_{\bk^{\prime}} \nonumber.
\end{equation}
The interaction potential is $W_{\bk, \bk^{\prime}} = V_{\bk, \bk^{\prime}}^{s,d} + U f_{\bk, \bk^{\prime}}^2 \chi(\bk^{\prime} - \bk)$.
The form factor $f_{\bk, \bk^{\prime}} = (h_{\bk} + h_{\bk^{\prime}})/2$, where $h_{\bk}$ transforms as $k_x^2-k_y^2$
in the $k_x - k_y$ plane.
A simple choice is $h_k = \cos(2\phi_{\bk})$, where $\phi_{\bk}$ is the azimuthal angle of $\bk$.
Note that the nematic pairing potential is intrinsically anisotropic in the presence of the nemato-elastic coupling.
This anisotropy can be taken into account by dividing the momentum space into two regions (a) $q_z \geq q_{2d}$, and (b)
$q_z \leq q_{2d}$, and by working with asymptotic forms of $\chi$
in these two regions.
Therefore, the pairing potential can be broken in three parts
\begin{align}
\label{eq:S-W-kernel}
W_{\bk, \bk^{\prime}} &= V_{\bk, \bk^{\prime}}^{s,d} + U f_{\bk, \bk^{\prime}}^2 \chi(\bk^{\prime} - \bk)|_{q_z\geq q_{2d}}
\nonumber \\
&+ U f_{\bk, \bk^{\prime}}^2 \chi(\bk^{\prime} - \bk)|_{q_z \leq q_{2d}}
\end{align}
The asymptotic forms of $\chi$ in the two regions are described in the main text.
Note, the critical manifold is contained in the third term above, while the second term above involves non-critical modes.
Furthermore, we will assume that $V > U$, where $V$ is the strength of the non-nematic pairing potential
$V_{\bk, \bk^{\prime}}^{s,d}$. Physically this
implies that sufficiently far from the nematic QCP the pairing is dominated by the non-nematic term. As noted in the main text,
the opposite limit of $U > V$ is trivial, since if the nematic potential already dominates pairing far away from the QCP,
then, irrespective of the strength of the nemato-elastic coupling, it will invariably lead to large $T_c$ enhancement because
the dominant pairing potential grows (even if it stays finite) as the QCP is approached.


(a) \textit{s-wave superconductivity with a uniform gap and a cylindrical Fermi surface (FS)}.
We will assume that the non-nematic pairing potential is a constant with $V_{\bk, \bk^{\prime}}^{s} = V$.
Since Eq.~(\ref{eq:BCS}) of the main text is restricted to the Fermi surface,
$q_{2D} = 2 k_F | \sin(\frac{\phi_{\bk} - \phi_{\bk}^{\prime}}{2}) |$
and $\cos^2(2\phi_{\bq}) = \cos^2(\phi_{\bk} + \phi_{\bk}^{\prime})$. The FS integral turns into angular integrals
$\lambda = <W_{\bk, \bk^{\prime}}>$ where
\begin{align}
<f> &= \int_{0}^{2\pi} \frac{du dv}{(2\pi)^2} f(u, v),
\end{align}
and $u= \phi_{\bk} + \phi_{\bk}^{\prime}$ and $v = \phi_{\bk} - \phi_{\bk}^{\prime}$.
This mean value is to be estimated to lowest order in the parameter $r \leq 1$ which governs the nearness to the QCP (see Fig.~1,
main text).

The contribution from the second term of Eq.~(\ref{eq:S-W-kernel}) is given by
\[
U \langle \cos^2 u \cos^2 v \frac{1}{r + (1- \cos v)/2} \rangle \approx \frac{U}{2 \sqrt{r}}.
\]
Note, in the above estimation typical $q_z \sim \pi/c$, where $c$ is unit cell length along the $z$-direction, while typical
$q_{2d} \sim r^{1/2}$. Therefore, to leading order in $r$
the constraint $q_z \geq q_{2d}$ is automatically satisfied in the above estimation, even though the angular integrals are
performed freely.

The third term of Eq.~(\ref{eq:S-W-kernel}) involves momentum dependence along the $z$-direction, and therefore, the estimation
of its contribution to $\lambda(r)$ involves averaging along the length of the cylindrical Fermi surface. Anticipating that the
typical momentum transfer along $z$ is small compared to Fermi wavevector $k_F$ we can write
$\oint_{\rm FS} \rightarrow \int_0^{2\pi} \frac{d \phi}{2 \pi} \int_0^1 d (k_z/k_F)$. This implies that the contribution from the
third term of Eq.~(\ref{eq:S-W-kernel}) is given by
\begin{widetext}
\[
2U \langle \cos^2 u \cos^2 v \left| \sin \frac{v}{2} \right| \int_0^1 dx \frac{1}{r -r_0 + r_0 \cos^2 u + \sin^2 (v/2) + r_0 x^2}
\rangle \approx - \frac{U}{\pi} \left( \ln {\rm max}[r-r_0,r_0] + c_1 \right),
\]
\end{widetext}
where $c_1 = 8/3 - 2 \ln 2 \approx 1.28$. This leads to the equation
\begin{equation}
\lambda (r \geq r_0)/\nu_{\rm FS}
= V  + \frac{U}{2 \sqrt{r}} - \frac{U}{\pi} \left( \ln {\rm max}[r-r_0,r_0] + c_1 \right) \nonumber,
\end{equation}
which is equation~(\ref{eq:result}) of the main text.
Note, the leading $r$-dependence comes from the non-critical modes,
rather than the critical ones which have a rather limited volume in momentum-space.
The critical contribution gives only to a weak
logarithmic dependence which can be ignored to leading order in $r$.

It is clear from the above that there will be considerable $T_c$ enhancement close to the
nematic QCP defined by $r=r_0$ only if in this regime the nematic contribution
dominates. This, in turn, is possible only if the nemato-elastic coupling is weak enough such that
\begin{equation}
r_0 < (U/V)^2 \nonumber.
\end{equation}
This is the condition mentioned in equation~(\ref{eq:condition}) of the main text.

In the following we consider few other cases and we show explicitly that the structure of Eq.~(\ref{eq:result})
remains the same, only numerical
pre-factors change. This implies that the conclusion obtained in Eq.~(\ref{eq:condition}) is robust.

(b) \textit{s-wave superconductivity with higher order gap harmonics}. Keeping s-wave symmetry we can introduce anisotropy
in the gap function by considering higher order harmonics as $\Delta(\bk) = \Delta_0 + \sqrt{2}\Delta_4 \cos(4\phi_{\bk})$.
The second term of r.h.s is the normalized first higher order s-wave harmonic.
We proceed to project the gap equation onto each orthogonal polynomial to get the secular equation
\begin{align}
\begin{split}
\lambda \Delta_0 &= \lambda_{00} \Delta_0 + \lambda_{40} \Delta_4, \\
\lambda \Delta_4 &= \lambda_{40} \Delta_0 + \lambda_{44} \Delta_4,
\end{split}
\end{align}
where we have defined $\lambda_{nn'} = <W_{kk'} g_n g_{n'}>$, with $g_n$ the n-th orthogonal cosine polynomial.
The secular system implies that the physical $T_c$ is to be given by the largest value of the matrix $(\lambda_{nn'})$.
With the above ansatz for the gap we get
\begin{equation}
\lambda = \frac{\lambda_{00} + \lambda_{44}}{2} + \sqrt{(\frac{\lambda_{00} - \lambda_{44}}{2})^2 + \lambda_{40}^2}.
\end{equation}
The calculation is then identical to case (a). Ignoring the log corrections from the critical manifold, we find to lowest order in $r$
\begin{align}
\lambda(r \geq r_0)/\nu_{FS} &= V + (1 + \frac{1}{\sqrt{2}})\frac{U}{2 \sqrt{r}},
\end{align}
which is same as in case (a) except for a numerical pre-factor.
It is also possible to do the calculation in the limit of an infinite number of s-wave harmonics, and
we find that the superconducting eigenvalue goes as $\lambda(r) = V + U/\sqrt{r}$.

(c) \textit{d-wave superconductivity}. Motivated by the cuprates, we take a non-nematic pairing interaction
which promotes d-wave superconductivity $V_{\bk, \bk^{\prime}}^{d} = 2 V \cos(2\phi_{\bk})\cos(2\phi_{\bk}^{\prime})$
on top of the nematic pairing potential.
With a $d$-wave gap ansatz $\Delta(\bk) = \Delta_0 \cos(2\phi_{\bk})$, we find
\begin{equation}
\lambda(r \geq r_0)/\nu_{FS} = V + \frac{3 U}{4 \sqrt{r}}.
\end{equation}
Thus, once again the BCS eigenvalue is the same as in Eq.~(\ref{eq:result}) except for a numerical pre-factor.

(d) \textit{The multiband case of Fe-based superconductors}.
Motivated by the physics of the Fe-based superconductors, we now consider a three band model with
one hole band centered around the $(0,0)$ point of the Brillouin zone and two electron pockets located at
$(\pi, 0)$ and $(0, \pi)$ respectively,
in the one-Fe/unit cell representation.
The non-nematic pairing potential is now a matrix in the band space, and we take it to be
\begin{equation}
V_{\bk, \bk^{\prime}}^{s} = -V \left( \begin{array}{ccc}
0 & 1/2 & 1/2 \\
1/2 & 0 & -1/2 \\
1/2 & -1/2 & 0 \\ \end{array} \right).
\end{equation}
Note, Eq.~(\ref{eq:BCS}) is written with the convention that repulsive interactions have negative sign and
attractive ones have positive sign. Thus, the above interaction implies that the non-nematic pairing potential is
only inter-band, and that it is repulsive for the electron-hole pairing term, while it is attractive for the
electron-electron pairing term. This invariably leads to a $s_{\pm}$ gap, which in the three-band language has the
form $\Delta_0 (1, -1, -1)$, which is the most discussed gap structure for these systems.
Note, the nematic pairing potential is attractive, and is, by definition, intra-band. For circular Fermi surfaces,
and assuming that the gaps on each of the pockets are constant, we get, following the case (a)
\begin{equation}
W_{\bk, \bk^{\prime}} = V \left( \begin{array}{ccc}
x & -1/2 & -1/2 \\
-1/2 & 2x & 1/2 \\
-1/2 & 1/2 & 2x \\ \end{array} \right),
\end{equation}
where $x= U/(2 \sqrt{r} V)$. This leads to a BCS eigenvalue
\begin{equation}
\lambda(r \geq r_0)/\nu_{FS} = \frac{1}{4}\left( V + \frac{3U}{\sqrt{r}} + \sqrt{ 9 V^2 + \frac{U^2}{r} + \frac{2 U V}{\sqrt{r}}}
\right).
\end{equation}
Since the only energy scales here are $V$ and $U/\sqrt{r}$, it is simple to check that significant $T_c$ enhancement is only
possible if the condition in Eq.~(\ref{eq:condition}) holds. Note also that, while the magnitudes of the gaps become different
on the hole and the electron pockets with the inclusion of the nematic pairing, both for small and for large $x$ there is a change
in the sign of the gap between the hole and the electron surfaces.

We conclude that in all the above cases the condition for significant $T_c$ enhancement is given by Eq.~(\ref{eq:condition}),
while in the opposite limit there is hardly any impact of the QCP on the $T_c$.

\end{document}